# Quantum Hall Effect in Electron-Doped Black Phosphorus Field-Effect Transistors

Fangyuan Yang[1,4,5], Zuocheng Zhang[1,4,5]*, Nai Zhou Wang[2,3,4], Guo Jun Ye[2,3,4], Wenkai Lou[6,7], Xiaoying Zhou[6,7], Kenji Watanabe[8], Takashi Taniguchi[8], Kai Chang[6,7], Xian Hui Chen[2,3,4]* and Yuanbo Zhang[1,4,5]*

[1]State Key Laboratory of Surface Physics and Department of Physics, Fudan University, Shanghai 200433, China
[2]Hefei National Laboratory for Physical Science at Microscale and Department of Physics, University of Science and Technology of China, Hefei, Anhui 230026, China
[3]Key Laboratory of Strongly Coupled Quantum Matter Physics, University of Science and Technology of China, Hefei, Anhui 230026, China
[4]Collaborative Innovation Center of Advanced Microstructures, Nanjing 210093, China
[5]Institute for Nanoelectronic Devices and Quantum Computing, Fudan University, Shanghai 200433, China
[6]SKLSM, Institute of Semiconductors, Chinese Academy of Sciences, PO Box 912, Beijing 100083, China
[7]Synergetic Innovation Center of Quantum Information and Quantum Physics, University of Science and Technology of China, Hefei, Anhui 230026, China
[8]National Institute for Materials Science, 1-1 Namiki, Tsukuba, 305-0044, Japan

**ABSTRACT:** The advent of black phosphorus field-effect transistors (FETs) has brought new possibilities in the study of two-dimensional (2D) electron systems. In a black phosphorus FET, the gate induces highly anisotropic 2D electron and hole gases. Although the 2D hole gas in black phosphorus has reached high carrier mobilities that led to the observation of the integer quantum Hall effect, the improvement in the sample quality of the 2D electron gas (2DEG) has however been only moderate; quantum Hall effect remained elusive. Here, we obtain high quality black phosphorus 2DEG by defining the 2DEG region with a prepatterned graphite local gate. The graphite local gate screens the impurity potential in the 2DEG. More importantly, it electrostatically defines the edge of the 2DEG, which facilitates the formation of well-defined edge channels in the quantum Hall regime. The improvements enable us to observe precisely quantized Hall plateaus in electron-doped black phosphorus FET. Magneto-transport measurements under high magnetic fields further revealed a large effective mass and an enhanced Landé g-factor, which points to strong electron-electron interaction in black phosphorus 2DEG. Such strong interaction may lead to exotic many-body quantum states in the fractional quantum Hall regime.

Black phosphorus has recently emerged as a new elemental two-dimensional (2D) semiconductor with high carrier mobilities.[1–5] The bandgap in few-layer black phosphorus varies with layer number,[6–10] strain[11–14] and electric field,[15,16] and falls in the spectral range from red to far-infrared.[17,18] The tunable bandgap, coupled with high sample quality,[19,20] makes black phosphorus a promising candidate for future electronic/optoelectronic applications. Field-effect transistors (FETs) based on black phosphorus exhibit ambipolar behavior.[1–5] A gate electric field induces a two-dimensional hole gas (2DHG) or electron gas (2DEG),[21–25] both of which are distinctively different from the 2D gases in conventional semiconductor heterostructures[26] or in graphene.[27,28] In particular, the new 2DHG and 2DEG exhibit peculiar electronic properties, such as high electronic anisotropy[4,5] and heavy carrier mass,[21–24] and greatly enrich the variety of 2D electron systems. Recent advances in device fabrication techniques have led to significant improvement in the quality of black phosphorus 2DHG, which culminated in the recent observation of quantum Hall effect (QHE).[19] The quantized Hall plateaus in black phosphorus 2DEG, however, have remained elusive due to the modest quality of black phosphorus 2DEG. Further in-depth studies of the 2DEG require high sample quality on the electron side of the gate doping. In particular, a high quality 2DEG on the electron side is crucial for future exploration of many-body physics in 2D; the higher effective mass in black phosphorus 2DEG implies a more prominent electron-electron interaction.

In this work, we observed for the first time quantized Hall plateaus in high-quality black phosphorus 2DEG under high magnetic fields. We achieved high quality in black phosphorus 2DEG by adopting a new device structure that was inspired by ref 29. Specifically, we used prepatterned graphite as a local gate to define the 2DEG in a selected region inside the black phosphorus flake; the area surrounding the 2DEG was in the meantime tuned into intrinsic insulating state with a Si back gate. The electrostatically defined smooth edge of the 2DEG is conducive to improved edge transport in the quantum Hall regime, and we observed well-developed Hall plateaus, accompanied by vanishing magnetoresistance under high magnetic fields. The transport study in the quantum Hall regime enabled us to delineate fundamental properties, the effective mass and Landé g-factor, of the carriers in black phosphorus 2DEG.

Specifically, the effective mass of the electron carriers is determined to be $m^* = (0.42 \pm 0.02)m_0$, where $m_0$ is the bare electron mass. Such large effective mass implies strong electron-electron interactions in black phosphorus 2DEG. Indeed, we obtained a much-enhanced g-factor of $5.7 \pm 0.7$ at the lowest filling factor of $\nu = 1$ (the g-factor at $\nu > 7$ reduces to $2.5 \pm 0.1$). Our results uncover the important role of electron-electron interaction in quantum transport through black phosphorus 2DEG, which may lead to new many-body states in the fractional quantum Hall regime.

We construct the black phosphorus FETs by sequentially stacking black phosphorus, hexagonal boron nitride (hBN) and graphite flakes. We start with a thin graphite flake (typical thickness ~ 10 nm) exfoliated on $SiO_2$/Si wafer. The graphite flake is then patterned into van der Pauw geometry (Figure 1a) with $O_2$ plasma reactive-ion etching, followed by annealing at 400 °C in forming gas to clean its surface. Meanwhile, we prepare black phosphorus/hBN stack by sequentially picking up black phosphorus and hBN flakes with a polypropylene carbon (PPC) film.[30] The black phosphorus/hBN stack is then released onto the prepatterned graphite flake, which serves as a local gate in addition to the Si back gate underneath (Figure 1a). We define Cr/Au contacts (2 nm Cr and 60 nm Au) on black phosphorus with standard e-beam lithography. Finally, we cover the device with another layer of hBN (thickness ~ 10 nm) to protect it from degradation in air. Because the quality of the 2D electron systems is largely determined by the cleanness of the black phosphorus/hBN interface, we took extra precautions to minimize defects/impurities as well as macroscopic bubbles introduced by the transfer processes. All transfer processes were performed in a glovebox with $O_2$ and $H_2O$ contents below 1 ppm, and we choose freshly cleaved black phosphorus and hBN flakes for device assembly. Figure 1b displays an optical image of the device (Device A), where the graphite local gate is outlined by broken lines. The metallic graphite local gate provides additional screening of charged impurities at the black phosphorus/hBN interface, which improves the quality of the 2D electron systems.[31,32] We obtain a Hall mobility of $\mu \sim 1300 \text{ cm}^{-2}\text{V}^{-1}\text{s}^{-1}$ at an electron density of $n = 2.7 \times 10^{12} \text{ cm}^{-2}$ at low temperature ($T = 0.3$ K; Supporting Information). Such a mobility is corroborated by the measurement of Shubnikov-de Haas (SdH) oscillations under magnetic fields; the oscillations start at an onset magnetic field of $B_c \approx 10$ T, which provides an independent estimation of the transport mobility $\mu \sim 1/B_c \approx 1000 \text{ cm}^{-2}\text{V}^{-1}\text{s}^{-1}$, in good agreement with the Hall mobility.

The prepatterned graphite gate enables us to define electrostatically the 2DEG in the flake, and we observe well-quantized Hall plateaus in black phosphorus 2DEG. A gate voltage $V_g$ applied on the graphite local gate induces free carriers in the region directly above the local gate, whereas the Si back gate (at voltage $V_{Si}$) depletes the rest of the surface. The two gates therefore completely determine the shape of the 2DEG, as well as its carrier density. More importantly, the 2DEG now has edges that are defined electrostatically by gate electric fields; the smooth edges are crucial for realizing quantized edge transport in the quantum Hall regime. Figure 1c,d displays the magnetoresistance $R_{xx}$ and Hall resistance $R_{xy}$, respectively, of Device A as a function of $V_g$ measured at four representative back gate voltages $V_{Si}$ under high magnetic field ($B = 35$ T) and low temperature ($T = 0.3$ K). QHE, which is characterized by quantized plateaus in Hall resistance $R_{xy}$ accompanied by vanishing magnetoresistance $R_{xx}$, is observed in black phosphorus 2DEG in all our data sets. The observation highlights the importance of gate-defined edge in the QHE regime, considering the fact that the mobility of Device A is not in fact the highest among reported values.[25,33] Close examination reveals that the QHE is most well-developed at around $V_{Si} = 5$ V, where the Si back gate depletes the black phosphorus around the 2DEG region, and creates well-defined boundaries for the QHE edge channels to propagate. As $V_{Si}$ deviates from the charge neutral point ($V_{Si} < 2$ V or $V_{Si} > 7$ V; Supporting Information), $R_{xx}$ at filling factors $\nu = 1$ and 2 becomes finite, probably due to dissipative conduction through the region that is not fully depleted. Finally, we note that effect of dissipative conduction is also visible in the Hall plateaus shown in Figure 1d. The plateau deviates from the quantized value at $\nu = 1$, when the black phosphorus around the 2DEG region becomes slightly electron-doped at $V_{Si} = 8$ V. For above reasons, the QHE data we present next are all taken at $V_{Si} = 5$ V. The observation of QHE in the electron-doped black phosphorus enables us to realize ambipolar quantum Hall effect in a single black phosphorus field-effect transistor (Supporting Information). The result sets the stage for future exploration of the ambipolar quantum edge transport in black phosphorus.

The well-developed QHE enables us to determine the energetics of the lowest Landau levels (LLs) in black phosphorus 2DEG. The fully-developed QHE at even and odd filling-factors (Figure 1c,d) indicates that all LLs are spin-split, and that the spin-split LLs are well separated by (generally filling-factor-dependent) energy gaps, $\Delta E_\nu$. The energy gaps are determined by two energy scales: Zeeman energy $E_Z = g^* \mu_B B$ and cyclotron energy $E_C = \hbar e B / m^*$, where $g^*$ is the effective Landé g-factor, $\mu_B$ is the Bohr magneton, $\hbar$ is the reduced Planck constant, $m^*$ is the effective mass of electrons and $B$ is the perpendicular magnetic field, respectively. Assuming a uniform LL broadening of $\Gamma$, $\Delta E_{odd} = E_Z - 2\Gamma$ and $\Delta E_{even} = E_C - E_Z - 2\Gamma$ (ref 34; here we have ignored the difference in $g^*$ induced by exchange interaction at odd and even filling-factors). In particular, $\Delta E_1$ is determined by Zeeman energy at $\nu = 1$, and is therefore linearly dependent on $B$; the slope of the linear dependence provides a measurement of $g^*$ of the 2DEG. To this end, we extracted $\Delta E_1$ from the thermally activated behavior of the $R_{xx}$ minima, $R_{xx}^{min} \sim e^{-\frac{\Delta E_1}{2k_B T}}$. Specifically, we obtain Arrhenius plots of $R_{xx}^{min}$ at $\nu = 1$ at various magnetic fields (Figure 2b) from temperature-dependent $R_{xx}$ as a function of $V_g$ displayed in Figure 2a. Line fits of Arrhenius plots yield the energy gap $\Delta E_1$ as a function of magnetic field as shown in Figure 2c. The linear relation between $\Delta E_1$ and magnetic field allows us to extract an effective Landé g-factor of $g_1 = 5.7 \pm 0.7$ at $\nu = 1$. Such a g-factor is significantly larger than the bare electron g-factor of $g = 2$ that is expected from the negligible spin-orbit coupling in black phosphorus.[35] Exchange interactions, however, may drastically enhance the g-factor in spin-polarized LLs.[36] Our results are clear experimental evidence of such many-body interactions in black phosphorus 2DEG.

We are now poised to delineate the energetics of higher LLs by analyzing the quantum oscillations in black phosphorus 2DEG. Despite the difficulty in obtaining the value of either $E_Z$ or $E_C$ from $E_\nu$ (mainly because of the large uncertainty in estimating $\Gamma$), the precise ratio between $E_Z$ and $E_C$ can be obtained through LL coincidence in a tilted magnetic field.[36,37] Measurement of $E_Z/E_C$ provides a precise determination of the magnetic susceptibility, $\chi = m^* g^*$, in unit of $\mu_B/2\pi\hbar^2$. We then perform a separate measurement of $m^*$ to unscramble $g^*$ from $\chi$ and thus completely determine $m^*$ and $g^*$, the two of the most fundamental parameters of an electron gas, at high filling factors.

We start with the measurement of $E_Z/E_C$. We performed the measurements in Device B (shown in Figure 3a inset; a graphite flake serves as back gate, and was not patterned) under a magnetic field of $B = 45$ T and a temperature of $T = 30$ mK. In a tilted magnetic field, Zeeman energy $E_Z = g^* \mu_B B_t$ and cyclotron energy $E_C = \hbar e B_\perp / m^*$, where $B_t$ and $B_\perp$ denote the total and perpendicular magnetic field respectively. As the magnetic field is titled at an angle $\theta$ away from the sample normal, $E_c$, which is proportional to $B_\perp = B_t cos\theta$, decreases, whereas $E_Z$, which is proportional to $B_t$, remains the same. When $cos\theta$ reaches a critical value $cos\theta_c = m^*g^*/2m_0$, $E_Z = E_C$, and the spin-split states from adjacent LLs coincide (such coincidence is referred to as the $r = 1$ coincidence discussed in ref 36). Consequently, energy gaps disappear at even filling-factors and the spin-split states merge into density of states (DOS) peaks. Further increase of $\theta$ resplits the DOS peaks and energy gaps reappear at even filling-factors. Such evolution of DOS at even filling-factors is sensitively captured by the amplitude of the SdH oscillations: The initial dips in $R_{xx}$ at even filling-factors evolve into peaks as $\theta$ increases, before re-emerging as shoulders at large $\theta$ (Figure 3a); $R_{xx}$ at even filling-factors reaches a maximum exactly at the coincidence angle $\theta_c$ (Figure 3b). It is therefore possible to obtain a precise measurement of $\theta_c$ (and thus $\chi = 2m_0 cos\theta_c$) through Gaussian fit of the $R_{xx}$ maxima,[36] as shown in Figure 3b (solid lines). The fit to all four data sets (at $\nu = 4, 6, 8$ and $10$) yields a mean critical angle of $\theta_c = 59.0°$. Such a coincidence angle yields a large magnetic susceptibility of $\chi = m^*g^* = 1.03m_0$, in the unit of $\mu_B/2\pi\hbar^2$ as shown in the inset of Figure 3b. Finally, we note that variations in the coincidence condition, if at all, were not apparent at different filling-factors, so the filling-factor-dependence of $\chi$ is not appreciable.

We now turn to the measurement of the effective mass $m^*$ in black phosphorus 2DEG. In conventional 2DEG in GaAs/AlGaAs heterostructures, the $m^*$ is normally obtained by fitting the temperature-dependent SdH oscillation amplitude $\Delta R_{xx}$ with Lifshitz-Kosevich (LK) formula[38]

$$\Delta R_{xx} \propto \lambda(T)/sinh\lambda(T) \quad (1)$$

where $\lambda(T) = 2\pi^2 k_B T m^*/\hbar eB$. Here $k_B$ is Boltzmann constant. The large Zeeman energy, however, poses a challenge in the measurement of $m^*$ in black phosphorus 2DEG: spin-splitting introduces a second harmonic into SdH oscillations, and Formula (1) breaks down.[38] We address the challenge by measuring $m^*$ in a tilted magnetic field:[37] near the coincidence condition, the spin-split states from adjacent LLs merge and the validity of Formula (1) is restored. Figure 4a displays the SdH oscillations at various temperatures when the field is tilted at $\theta = 48.5°$, where the adjacent LLs merge (Supporting Information). The structure of Device C is the same as Device B. Although such tilt angle is not exactly the coincidence angle, the fast Fourier transform (FFT) of the SdH oscillations (Figure 4a inset) indicates that the oscillations are now dominated by the first harmonic and the second harmonic is negligible. By fitting the temperature-dependent oscillation amplitude with Formula (1) at different filling factors (Figure 4b), we obtained an effective mass of $m^* = (0.42 \pm 0.02)m_0$, shown in the upper panel of Figure 4c. We note that $m^*$ is weakly dependent on the filling-factor, which may be attributed to the fact that at a tilt angle the in-plane magnetic field distorts the Fermi surface.[39] The result provides a more precise determination of the effective mass than a previous attempt that included the second harmonic in the fitting.[22] It also agrees reasonably well to theoretical values reported in the literature,[22,35] which ranges from $0.42m_0$ to $0.45m_0$ within a thickness of 1 to 5 layers. Above measurements of $\chi$ and $m^*$ yield a $g^* = 2.5 \pm 0.1$ at high LLs (Figure 4c, lower panel).

Our analysis of $g^*$ and $m^*$ provides valuable insights into the electronic structure of black phosphorus 2DEG at the extreme quantum limit. There are two main points to note. First, the much-enhanced g-factor of $g_1 = 5.7 \pm 0.7$ at $\nu = 1$ points to the predominant role of electron-electron interactions in the LL energetics at the lowest filling factors. The g-factor reduces to $g^* = 2.5 \pm 0.1$ at higher filling factors ($\nu > 7$), which we attribute to the strong screening effect at high electron densities that suppresses electron-electron interactions. Meanwhile, increased LL broadening at high filling factors may reduce the spin polarization, which also leads to a g-factor close to the bare electron g-factor of $g = 2$. Second, the large effective mass of the carriers suppresses the kinetic energy (relative to the Coulomb energy), and underpins the strong electron-electron interaction in black phosphorus 2DEG. The large $g^*$ and $m^*$ yield a magnetic susceptibility that reaches $2.4m_0$ at $\nu = 1$. The value is on the same order of magnitude with that in the 2DEG in ZnO/MgZnO heterostructures[40] and AlAs quantum well[41] but is significantly larger than that in GaAs/AlGaAs heterostructure ($0.1m_0$; ref 42). Consequently, Zeeman energy becomes a defining energy scale in black phosphorus 2DEG under magnetic fields, which may lead to the observation of exotic even-denominator fractional quantum Hall effect.[40]

In conclusion, we fabricated black phosphorus FETs with prepatterned graphite local gate that electrostatically defines the 2DEG and its edge. Such a device structure enabled us to observe well-quantized Hall plateaus on electron-doped black phosphorus FETs. Magneto-transport measurements in the quantum Hall regime revealed a much-enhanced Landé g-factor at the lowest filling factor, providing unambiguous evidence of strong electron-electron interactions in black phosphorus 2DEG. Indeed, we obtained a large effective carrier mass that underlies the strong interaction effects. The strong electron-electron interaction, coupled with the anisotropic electronic structure of black phosphorus, may lead to exotic ground states in the fractional quantum Hall regime.

## ASSOCIATED CONTENT

**Supporting Information.**

Characterization of carrier mobility in Device A;

Magnetoresistance of Device A under various back-gate biases;

Ambipolar quantum Hall effect in Device B;

Quantum oscillations in Device C in tilted magnetic fields

## AUTHOR INFORMATION

**Corresponding Author**

*E-mail: zuocheng.zhang@fudan.edu.cn

*E-mail: chenxh@ustc.edu.cn

*E-mail: zhyb@fudan.edu.cn

**Author Contributions**

Z.Z., X.H.C., and Y.Z. cosupervised the project. F.Y. and Z.Z. fabricated the device and carried out transport measurements. N.Z.W. and G.J.Y. grew bulk crystal black phosphorus. W.L., X.Z, and K.C. helped with data analysis. K. W. and T. T. grew hBN crystal. F.Y., Z.Z., and Y.Z. wrote the paper with input from all authors.

**Notes**

The authors declare no competing interests.

## ACKNOWLEDGMENT

A portion of this work was performed at the National High Magnetic Field Laboratory, which is supported by National Science Foundation Cooperative Agreement No. DMR-1157490 and DMR-1644779 and the State of Florida. Part of the sample fabrication was conducted at Fudan Nanofabrication Lab. F.Y., Z.Z. and Y.Z. acknowledge financial support from National Key Research Program of China (Grants 2016YFA0300703 and 2018YFA0305600), and NSF of China (Grants U1732274, 11527805, 11425415, 11421404 and 11604053). Z.Z. also acknowledges support from China Postdoctoral Science Foundation (Grants 2016M600279 and 2017T100264). N.Z.W, G.J.Y and X.H.C. acknowledge support from the 'Strategic Priority Research Program' of the Chinese Academy of Sciences (Grants XDB04040100) and the National Basic Research Program of China (973 Program; Grant 2012CB922002). X.H.C. also acknowledges support from the National Natural Science Foundation of China (Grant 11534010) and the Key Research Program of Frontier Sciences, CAS (Grant QYZDY-SSW-SLH021). W.L, X.Z. and K.C. acknowledge support from NSF of China (Grant 11434010). K.W. and T.T. acknowledge support from the Elemental Strategy Initiative conducted by the MEXT, Japan and JSPS KAKENHI (Grant JP15K21722).

# FIGURE CAPTIONS:

**Figure 1.** Device structure and the observation of QHE in electron-doped black phosphorus FET. (a,b) Schematic device structure (a) and optical image (b) of black phosphorus FET with prepatterned graphite local gate and Si back gate. Graphite local gate induces a 2DEG in black phosphorus directly above the local gate, whereas Si back gate depletes the rest of the black phosphorus surface. The edge of the 2DEG is therefore electrostatically defined by the two gates. (c,d) Magnetoresistance $R_{xx}$ (c) and Hall resistance $R_{xy}$ (d) recorded as functions of local-gate voltage $V_g$ at $B = 35$ T and $T = 0.3$ K at representative back-gate biases. QHE, characterized by quantized plateaus in $R_{xy}$ and vanishing $R_{xx}$, is observed at all back-gate biases. The QHE is most well-developed at back-gate voltages around $V_{Si} = 5$ V that corresponds to the charge neutral point of the black phosphorus FET.

**Figure 2.** Measurement of energy gaps in the QHE regime. (a) $R_{xx}$ as a function of $V_g$ recorded at various temperatures. Data were collected under a magnetic field of $B = 35$ T and a back-gate bias of $V_{Si} = 5$ V. (b) Arrhenius plots of the minimum resistance $R_{xx}^{min}$ at $\nu = 1$ at various magnetic fields. For each magnetic field, the line fit to $lnR_{xx}^{min}$ as a function of $1/T$ yields the energy gap at $\nu = 1$. (c) Energy gap extracted in (b) plotted as a function of magnetic field. Line fit of the data set gives an estimation of effective g-factor at $5.7 \pm 0.7$ at $\nu = 1$.

**Figure 3.** Quantum oscillations in tilted magnetic field. (a) Magnetoresistance as a function of $(V_g - V_0)/cos\theta$ under tilted magnetic fields. Here $V_0$ is the charge neutral point of the device, and $\theta$ is the tilt angle. Upper inset displays the optical image of the device (Device B). Lower inset illustrates the measurement configuration and a diagram of the LLs under coincidence condition. Data were recorded at $T = 30$ mK with the total magnetic field fixed at $B_t = 45$ T. Magnetoresistance curves are shifted by 1 k$\Omega$ for clarity. Broken lines mark the position of the Zeeman gap, which are labeled by odd filling-factors. (b) Resistance maxima at even filling-factors ($\nu = 4, 6, 8, 10$) measured in the neighborhood of the critical angle where LL coincidence takes place. Solid lines are Gaussian fits to the four data sets. Inset: $g^*m^*/m_0 = 2cos\theta_c$ as a function of filling-factor obtained from the Gaussian fit. The dotted line indicates the mean value of $g^*m^*/m_0$.

**Figure. 4.** Temperature-dependent SdH oscillations. (a) Oscillatory part of $R_{xx}$ plotted as a function of $1/Bcos\theta$ at various temperatures. The measurements were performed at a fixed tilt angle $\theta = 48.5°$ near the LL coincidence in Device C.

Integers mark the filling-factors. Inset: FFT of the SdH oscillations shown in the main figure. (b) SdH oscillation amplitude extracted from (a) plotted as a function of temperature. Solid lines are fits with LK formula described in the main text. (c) Upper panel: effective mass $m^*$ of electron carriers (normalized by $m_0$) extracted from fits in (b). Lower panel: effective Landé g-factor $g^*$ of electron carriers obtained from measurements of $\chi$ and $m^*$. Broken lines mark the mean value of $m^*/m_0$ and $g^*$ in the upper and low panel, respectively.

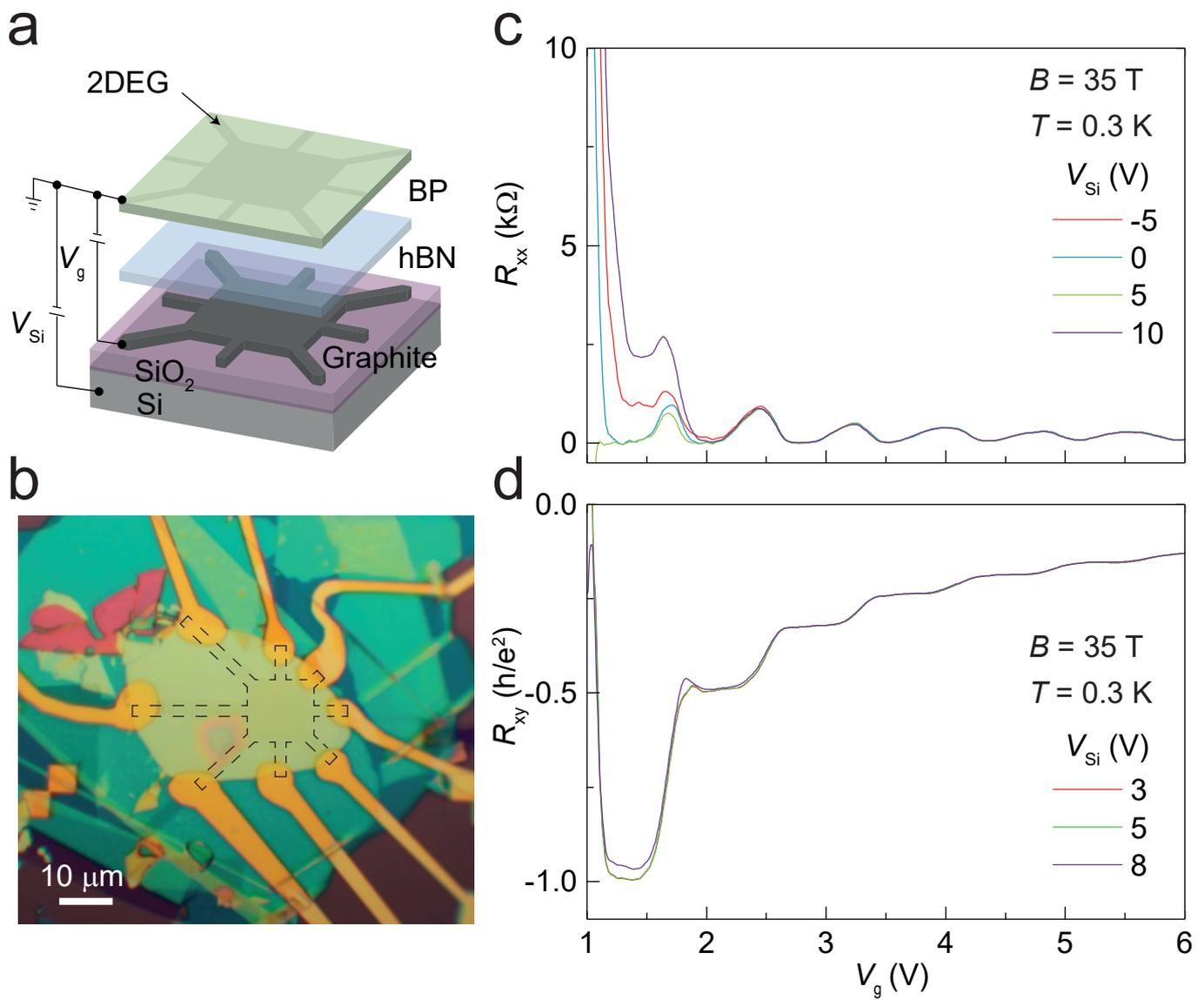

Figure 1

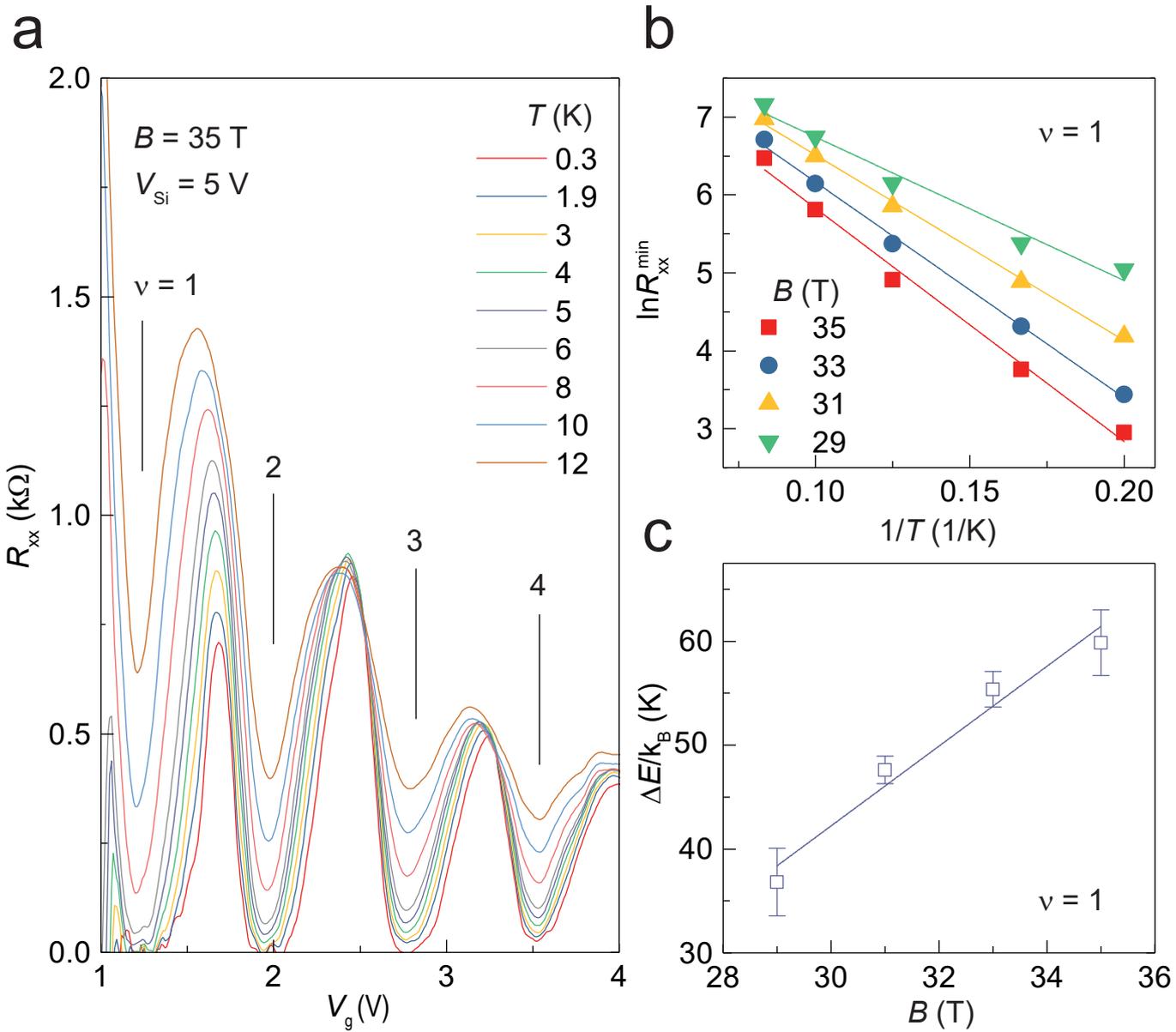

Figure 2

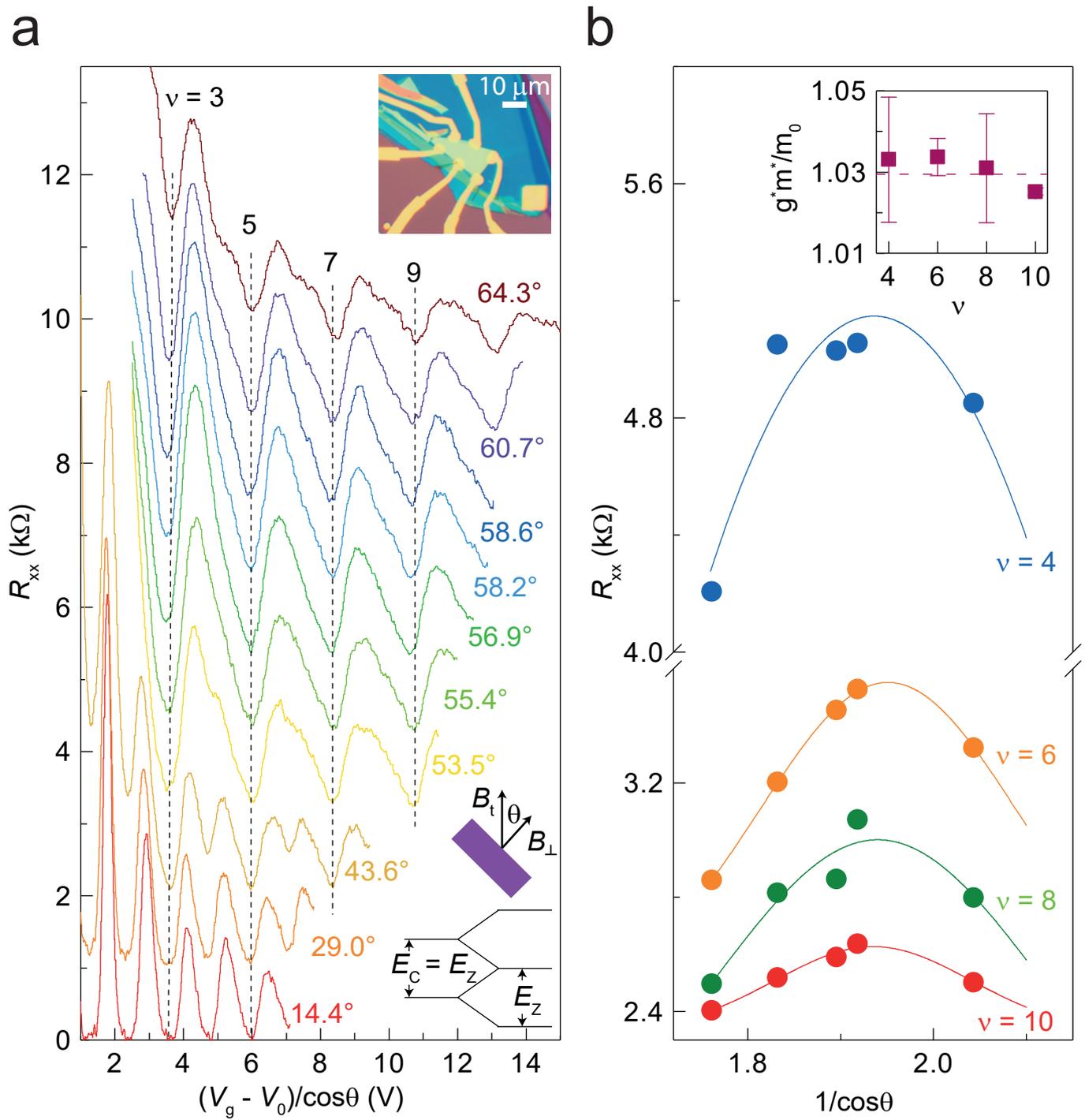

Figure 3

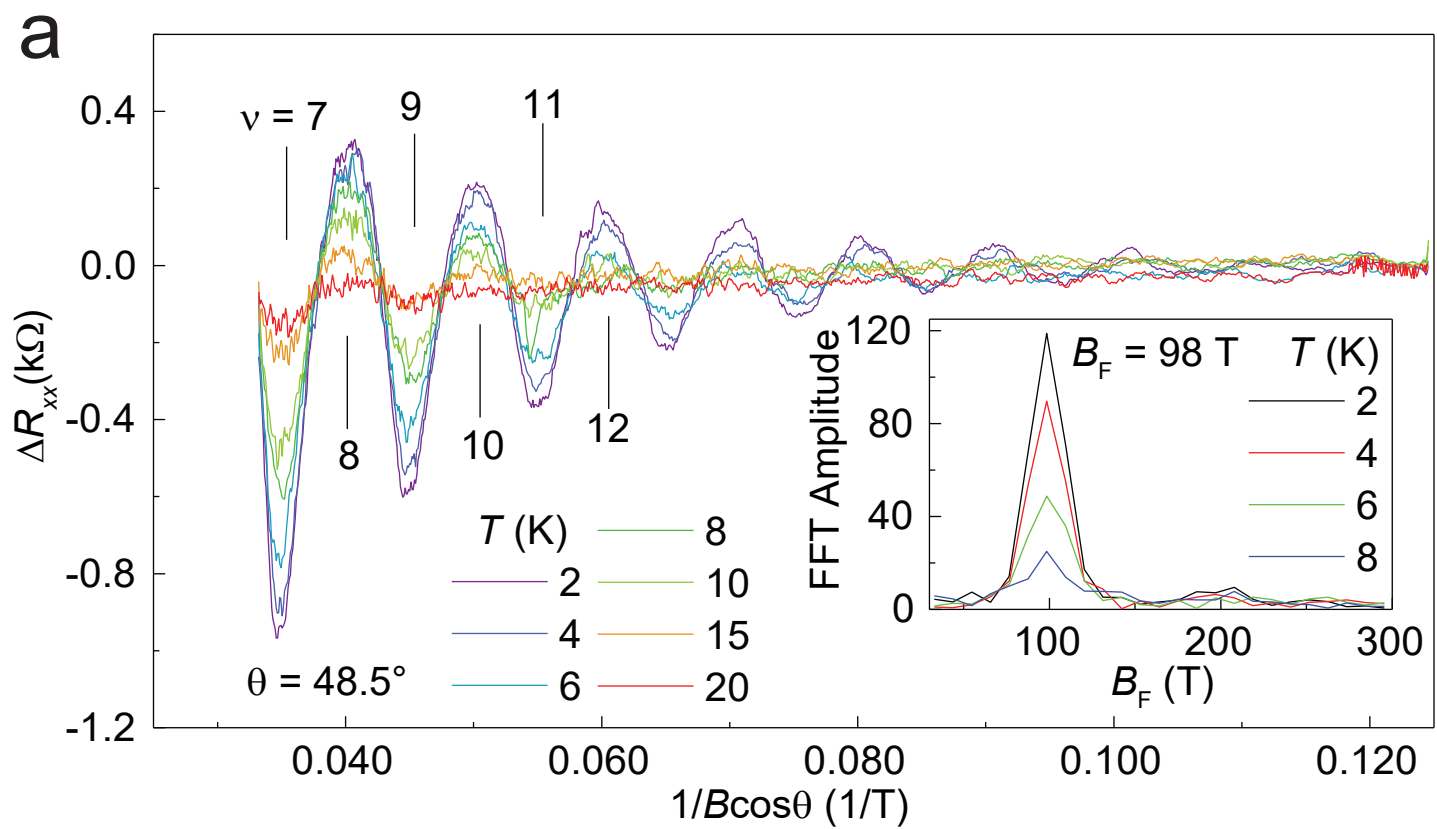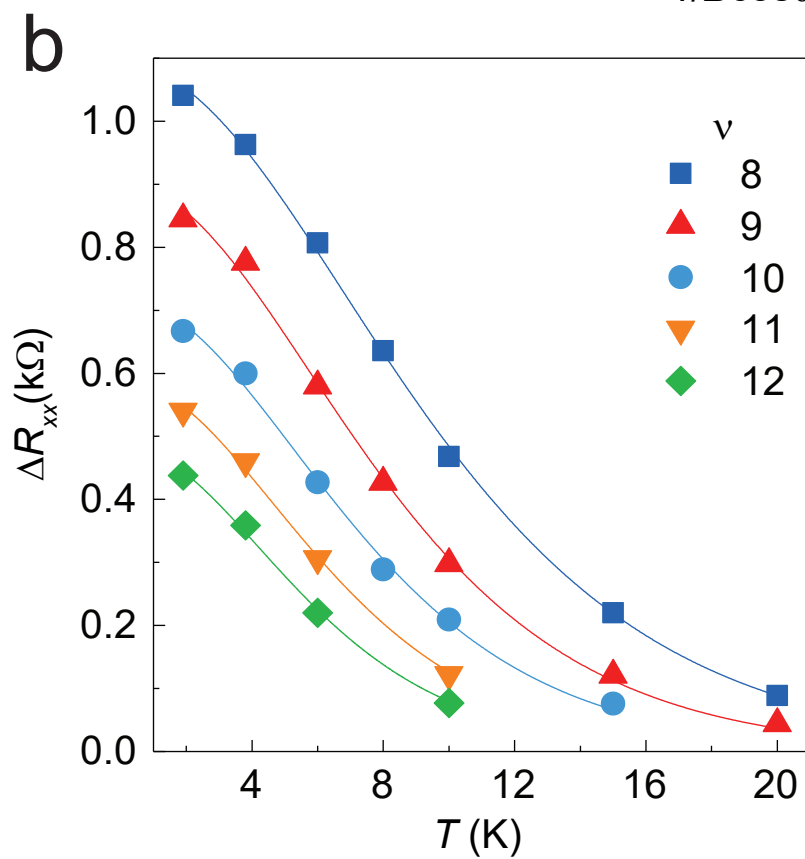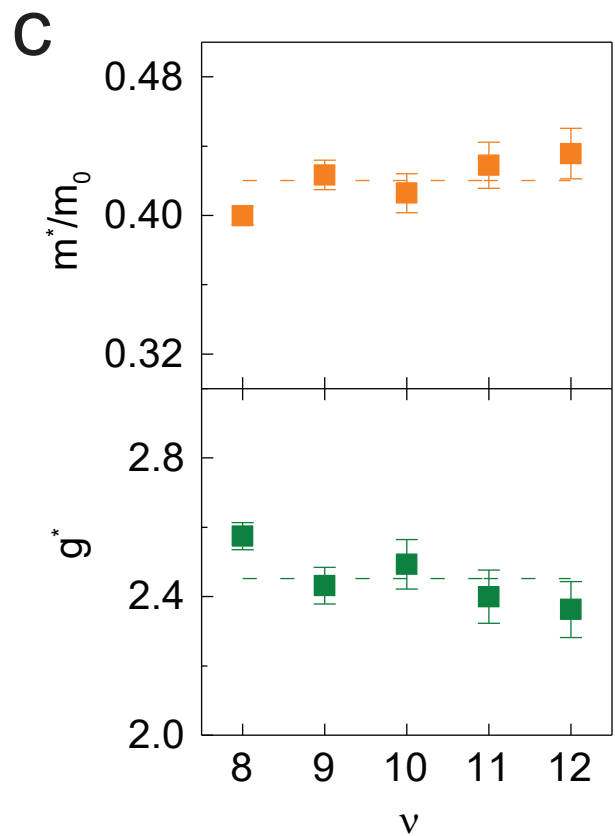

Figure 4